\documentclass[aps,10pt,prb,twocolumn,superscriptaddress]{revtex4-2}
\usepackage{graphicx}
\usepackage{amsmath}
\usepackage{amsfonts,hyperref}
\usepackage{amssymb,mathrsfs}
\usepackage{upgreek}
\usepackage{bbold,yfonts}

\newcommand{\bs}[1]{\boldsymbol{#1}}

\begin{document}

\title{Corbino magnetoresistance in neutral graphene}

\author{Vanessa Gall} \affiliation{\mbox{Institute for Quantum
    Materials and Technologies, Karlsruhe Institute of Technology,
    76021 Karlsruhe, Germany}} \affiliation{\mbox{Institut f\"ur
    Theorie der Kondensierten Materie, Karlsruhe Institute of
    Technology, 76128 Karlsruhe, Germany}}

\author{Boris N. Narozhny} \affiliation{\mbox{Institut f\"ur Theorie
    der Kondensierten Materie, Karlsruhe Institute of Technology,
    76128 Karlsruhe, Germany}} \affiliation{National Research Nuclear
  University MEPhI (Moscow Engineering Physics Institute), 115409
  Moscow, Russia}

\author{Igor V. Gornyi} \affiliation{\mbox{Institute for Quantum
    Materials and Technologies, Karlsruhe Institute of Technology,
    76021 Karlsruhe, Germany}} \affiliation{\mbox{Institut f\"ur
    Theorie der Kondensierten Materie, Karlsruhe Institute of
    Technology, 76128 Karlsruhe, Germany}} \affiliation{Ioffe
  Institute, 194021 St.~Petersburg, Russia}

\date{\today}

\begin{abstract}
  We explore the magnetohydrodynamics of Dirac fermions in neutral
  graphene in the Corbino geometry. Based on the fully consistent
  hydrodynamic description derived from a microscopic framework and
  taking into account all peculiarities of graphene-specific
  hydrodynamics, we report the results of a comprehensive study of the
  interplay of viscosity, disorder-induced scattering, recombination,
  energy relaxation, and interface-induced dissipation. In the clean
  limit, magnetoresistance of a Corbino sample is determined by
  viscosity.  Hence the Corbino geometry could be used to measure the
  viscosity coefficient in neutral graphene.
\end{abstract}

\maketitle

Transport measurements remain one of the most common experimental
tools in condensed matter physics. Having dramatically evolved past
the original task of establishing bulk material characteristics such
as electrical and thermal conductivities, modern experiments often
involve samples that are tailor-made to target particular properties
or behavior.

In recent years considerable efforts have been devoted to uncovering
the collective or hydrodynamic flows of charge carriers in ultraclean
materials as predicted theoretically \cite{pg,me0,luc,rev}. Several
dedicated experiments focused on answering two major questions: is the
observed electronic flow really hydrodynamic and how to measure
electronic viscosity \cite{geim1,geim4,imh,imm,geim2,young}, the
quantity that fascinates physicists beyond the traditional condensed
matter physics
\cite{stein19,han20,brad,gold18,kov,Schafer2014,zaa13,hema}.  The
hydrodynamic regime is apparently easiest to achieve in graphene
\cite{me0,luc,rev}. This material is especially interesting since it
can host two drastically different types of hydrodynamic behavior: (i)
``conventional'' at relatively high carrier densities
\cite{gurzhi,fl0,luc} and (ii) ``unconventional'' at charge neutrality
\cite{hydro1,me1}.

Linearity of the excitation spectrum in graphene leads to the fact
that electronic momentum density defines the energy current,
$\bs{j}_E$. In the intermediate temperature window where
electron-electron interaction is the dominant scattering process in
the system ($\ell_{ee}\ll\ell_{\rm dis},\ell_{e-ph},W$, in the
self-evident notation) the energy flow becomes hydrodynamic. At high
carrier densities (in ``doped graphene'') the energy current is
essentially equivalent to the electric current, $\bs{j}$, allowing one
to formulate a Navier-Stokes-like equation for $\bs{j}$ \cite{fl0} as
pioneered by Gurzhi \cite{gurzhi}.

At charge neutrality and in the absence of the external magnetic field
($\bs{B}=0$) the energy and electric currents decouple
\cite{hydro0}. In the hydrodynamic regime the electric current remains
Ohmic \cite{me1} (with the ``internal'' or ``quantum'' conductivity
$\sigma_Q$ due to electron-electron interaction
\cite{kash,mfss,mfs,schutt}), while the Navier-Stokes-like equation
describes the energy current \cite{me1,me3,megt}. If external magnetic
field is applied, the energy and charge flows become entangled
\cite{hydro0,me1,hydro1} allowing for a possibility to detect the
hydrodynamic flow in electronic transport experiments. In particular,
a bulk (infinite) system is characterized by positive, parabolic
magnetoresistance \cite{hydro0,mus} proportional to the disorder mean
free time $\tau_{\rm dis}$ (disorder scattering is the only mechanism
of momentum relaxation).

\begin{figure}[t]
\centering
\includegraphics[width=\columnwidth]{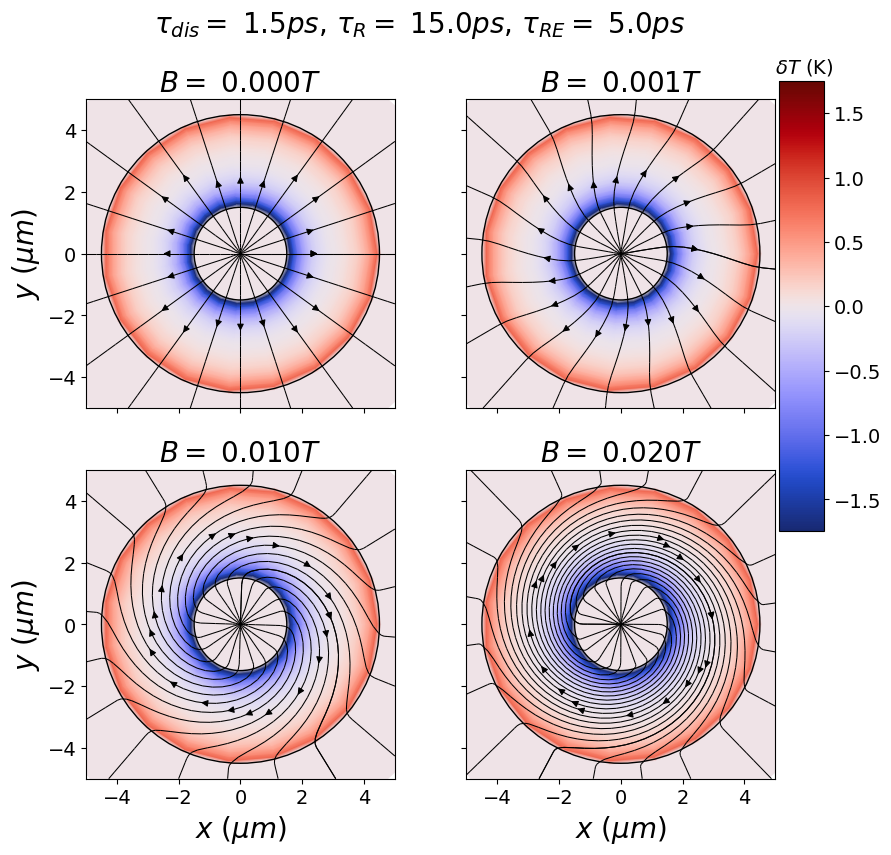}
\caption{Hydrodynamic velocity $\bs{u}$ and temperature $\delta T$
  distribution in the device obtained by solving the hydrodynamic
  equations at relatively high temperatures where energy relaxation is
  dominated by supercollisions. Arrows indicate $\bs{u}$ and the color
  map shows $\delta T$. The quantitative results were computed using
  the following values of the average temperature $T=150\,$K, disorder
  scattering time $\tau_{\rm dis}=1.5\,$ps (corresponding to the
  scattering rate
  $\tau^{-1}_{\rm{dis}}\approx0.67\,$THz$\,\approx5.1\,$K),
  recombination time $\tau_R=15\,$ps, energy relaxation time
  $\tau_{RE}=5\,$ps, dimensionless coupling constant in graphene
  $\alpha=0.5$, carrier density in the leads
  $n_L=5\times10^{12}\,$cm$^{-2}$, and the current passing through the
  device $I=1\,\mu$A. The four panels correspond to the indicated
  values of magnetic field.}
\label{fig:ut}
\end{figure}

The outcome of a given measurement is strongly influenced by the
sample size and geometry. Early experiments focused on either the
``strip'' (or Hall bar) \cite{geim1,geim4,imh,imm} or the point
contact geometry \cite{geim2,young}, while more recently data on
Corbino disks became available \cite{sulp22}.

The simplest viscous phenomenon one can look for in a long
(striplike) sample
\cite{mol95,imm,imh,vool,mac,geim5,geim2,young,gus18,gus20,gus21,rai20,gupta,var20,goo2,jao,mol95,pol15,ale,moo,pol17,mr2,cfl,ady2,han20,glazman20,sven1}
is the Poiseuille flow \cite{poi,poise,dau6}. This flow is
characterized by a parabolic velocity profile with the curvature
determined by viscosity. In doped graphene the Poiseuille flow of
charge can be detected by imaging the electric current density
\cite{imm}. In contrast, neutral graphene exhibits the Poiseuille flow
of the energy current \cite{julia}. Moreover, at relatively high
temperatures where hydrodynamic behavior in graphene is observed the
electron-phonon interaction (either direct \cite{hydro0,fos16,lev19}
or via ``supercollisions'' \cite{meig1,srl,ralph13,betz,tik18,kong})
cannot be neglected and hence electronic energy is not conserved. The
resulting energy relaxation dwarfs the viscous contribution to the
Navier-Stokes \cite{van1} equation.

Applying a perpendicular magnetic field to a neutral graphene strip
leads to a coupled charge and energy flow with the two currents being
orthogonal \cite{hydro0}. The electric current flowing along the strip
is accompanied by a neutral quasiparticle flow in the lateral
direction resulting in energy and quasiparticle accumulation near the
strip boundaries \cite{meg,mr1}. The accumulation is limited by
quasiparticle recombination \cite{mr1} and energy relaxation processes
\cite{meig1}. As a result, the boundary region's contribution to the
resistance is linear in the applied magnetic field
\cite{hydro0,mr1,mr3,mr2}, in contrast to the standard quadratic
magnetoresistance of the bulk system \cite{hydro0,mus}. In classically
strong fields the boundary contribution dominates making the linear
magnetoresistance directly observable. This effect is not specific to
Dirac fermions as shown by experiments in bilayer graphene
\cite{mrexp}.

The Corbino geometry presents an interesting alternative to the Hall
bar experiments
\cite{corbino,corb,ady,fal19,sulp22,oga21,alex22,ady22,rai22,alex22-2,van1}. In
a typical measurement the electric current is passed from the inner to
the outer boundary of a Corbino disk. The specific feature of the
stationary flow in this geometry is that the magnitude of the radial
component of the current is determined by the continuity equation
alone. In the absence of the magnetic field the whole current flows
radially. Combining the solution of the continuity equation with the
hydrodynamic Gurzhi equation (e.g., in doped graphene) leads to an
apparent paradox \cite{fal19}: the current flow appears unaffected by
viscosity. However, the dissipated energy is still determined by
viscosity leading to the jumps of electric potential at the contacts
thus resolving the paradox. In a perpendicular magnetic field the system
exhibits parabolic magnetoresistance inverse proportional to the
viscosity and independent of the disorder scattering. Applied
phenomenologically to neutral graphene (neglecting contact effects)
\cite{alex22-2} this conclusion stands in sharp contrast to the
standard result \cite{hydro0,mus} raising the question of the fate of
the disorder-limited bulk magnetoresistance in the Corbino geometry.

In this paper we investigate hydrodynamic flows in neutral graphene in
the Corbino disk subjected to the perpendicular magnetic field based
on the graphene-specific hydrodynamic theory \cite{me0,me1,meig1}
reporting the results of a careful study of the interplay of
viscosity, disorder-induced scattering, recombination, energy
relaxation, and interface-induced dissipation. Solving the
hydrodynamic equations we find the spatial distribution of the
hydrodynamic velocity $\bs{u}$, temperature (see Fig.~\ref{fig:ut}),
electric current, and potential $\varphi$ (see Fig.~\ref{fig:jf}).
Furthermore, we calculate the field-dependent resistance of the whole
Corbino sample including the leads. Keeping in mind recent and ongoing
experiments, it appears logical to include the effect of the lead
resistance in order to achieve a more realistic description of the
Corbino device. However, the theoretical limit of ``ideal'' leads can be
considered without any complications.

The main results of this paper are as follows. We show that
magnetoresistance of the Corbino device exhibits a crossover from the
``hydrodynamic'' (viscosity-dominated) to the ``bulk''
(disorder-limited) behavior with the increasing system size as
compared to the Gurzhi length ${\ell_G = \sqrt{\nu\tau_{\rm dis}}}$
\cite{moo,pol17,mr2,cfl,sven1} ($\nu$ is the kinematic viscosity
\cite{dau6,luc,geim1,geim4,me2} and $\tau_{\rm dis}$ is the disorder
mean free time). In the clean limit
(${\tau_{\rm{dis}}\rightarrow\infty}$) magnetoresistance remains
finite and is determined by viscosity offering a way to measure the
viscosity coefficient in neutral graphene. In classically strong fields
magnetoresistance remain parabolic (in contrast to the linear
magnetoresistance in the strip geometry). The ``contact
magnetoresistance'' induced through the dissipation jump is present,
but is typically weaker than the bulk contribution.

\section{Magnetohydrodynamics in graphene}

Our arguments are based on the hydrodynamic theory of electronic
transport in neutral graphene derived from the kinetic (Boltzmann)
equation \cite{hydro1,me1,meig1} or from the microscopic Keldysh
technique \cite{me23}. At charge neutrality both bands contribute to
transport on equal footing. A current-carrying state is characterized
by the chemical potentials $\mu_\pm$ of each band or by their linear
combinations \cite{alf,me1}
\begin{subequations}
\label{muns}
\begin{equation}
\label{mus}
\mu = \frac{\mu_+\!+\!\mu_-}{2},
\qquad
\mu_I = \frac{\mu_+\!-\!\mu_-}{2},
\end{equation}
conjugate to the ``charge'' and ``imbalance'' (or ``total quasiparticle'') densities
\begin{equation}
\label{ns}
n = n_+ - n_-, \qquad
n_I = n_+ + n_-.
\end{equation}
\end{subequations}
In equilibrium ${\mu_I=0}$. Any macroscopic current can be expressed
as a product of the corresponding density and hydrodynamic velocity
$\bs{u}$ (up to dissipative corrections). Due to the kinematic
peculiarity of the Dirac fermions in graphene known as the ``collinear
scattering singularity'' \cite{hydro1,mfss} one has to consider the
electric, energy, and imbalance, $\bs{j}_I$ currents defined as
\begin{equation}
\label{jlr}
\bs{j} = n\bs{u} \!+\! \delta\bs{j},
\quad
\bs{j}_I = n_I\bs{u} \!+\! \delta\bs{j}_I,
\quad
\bs{j}_E = {\cal W}\bs{u},
\end{equation}
where ${\cal W}$ is the enthalpy density and $\delta\bs{j}$ and
$\delta\bs{j}_I$ are the dissipative corrections. In the degenerate
limit ${\mu\gg{T}}$ the dissipative corrections vanish \cite{me1,me3}
justifying the applicability of the single-band picture to doped
graphene. At charge neutrality ${n=0}$, the electric and energy
currents in Eq.~(\ref{jlr}) appear to be decoupled \cite{me1}.

Within linear response, steady-state macroscopic currents obey the
linearized hydrodynamic equations \cite{megt2}. Assuming that the
dominant mechanism of energy relaxation is supercollisions
\cite{meig1}, the equations have the form
\begin{subequations}
\label{hydrolin1}
\begin{equation}
\label{cen2}
\bs{\nabla}\!\cdot\!\delta\bs{j} = 0,
\end{equation}
\begin{equation}
\label{ceni2}
n_{I}\bs{\nabla}\!\cdot\!\bs{u} + \bs{\nabla}\!\cdot\!\delta\bs{j}_I 
= -\frac{12\ln2}{\pi^2}\frac{n_{I}\mu_I}{T\tau_R},
\end{equation}
\begin{eqnarray}
\label{nseqlin1}
\bs{\nabla} \delta P
=
\eta \Delta\bs{u}
+
\frac{e}{c}\delta\bs{j}\!\times\!\bs{B}
-
\frac{3P\bs{u}}{v_g^2\tau_{{\rm dis}}},
\end{eqnarray}
\begin{equation}
\label{tteqlin1}
3P\bs{\nabla}\!\cdot\!\bs{u} = - \frac{2\delta{P}}{\tau_{RE}}.
\end{equation}
\end{subequations}
Here Eq.~(\ref{cen2}) is the continuity equation; Eq.~(\ref{ceni2}) is
the ``imbalance'' continuity equation \cite{alf,me1} (where $v_g$ is
the band velocity in graphene, $c$ is the speed of light, $e$ is the
unit charge, and $\tau_R$ is the recombination time);
Eq.~(\ref{nseqlin1}) is the linearized Navier-Stokes equation
\cite{msf,me1,megt,megt2} (with $\eta$ being the shear viscosity); and
Eq.~(\ref{tteqlin1}) is the linearized ``thermal transport'' equation
($\tau_{RE}$ is the energy relaxation time \cite{meig1}). We follow
the standard approach \cite{dau6} where the thermodynamic quantities
are replaced by the corresponding equilibrium functions of the
hydrodynamic variables. Equilibrium thermodynamic quantities, i.e.,
the pressure $P=3\zeta(3)T^3/(\pi v_g^2)$, enthalpy density
${\cal{W}}$, imbalance density, ${n_{I}=\pi{T}^2/(3v_g^2)}$, and
energy density are related by the ``equation of state'',
${\cal{W}}=3{P}=3{n}_E/2$. Equations (\ref{hydrolin1}) should be
solved for the unknowns $\bs{u}$, $\mu_I$, and $\delta P$ keeping the
remaining (thermodynamic) quantities, e.g., $n_{I}$, $P$, and $T$,
constant.

The dissipative corrections to the macroscopic currents can be
determined from the underlying microscopic theory
\cite{me1,megt,megt2} and are expressed in terms of the same variables
closing the set of hydrodynamic equations (\ref{hydrolin1})
\begin{subequations}
\label{djs}
\begin{equation}
\label{lj}
\delta\bs{j} = 
\frac{1}
{e^2\tilde{R}}
\left[ e\bs{E}
+
\omega_B\bs{e}_B\!\times\!
\left(
\frac{\alpha_1\delta_I\bs{\nabla}\mu_I}{\tau_{\rm dis}^{-1}\!+\!\delta_I^{-1}\tau_{22}^{-1}}
-
\frac{2T\ln2}{v_g^2}\bs{u}\right)\!\right]\!,
\end{equation}
\begin{eqnarray}
\label{lji}
&&
\!\!\!\!\!
\delta\bs{j}_I = 
-\frac{\delta_I}{\tau_{\rm dis}^{-1}\!+\!\delta_I^{-1}\tau_{22}^{-1}}
\frac{1}{e^2\tilde{R}}\times
\\
&&
\nonumber\\
&&
\;\;
\times\!
\left[ \alpha_1\omega_B\bs{e}_B\!\times\!\bs{E}
\!+\!
\frac{2T\ln2}{\pi}e^2R_0\bs{\nabla}\mu_I
\!+\!
\alpha_1\omega_B^2\frac{2T\ln2}{v_g^2}
\bs{u} \right]\!,
\nonumber
\end{eqnarray}
\begin{equation}
\tilde{R}=R_0\!+\!\alpha_1^2\delta_I\tilde{R}_B.
\end{equation}
\end{subequations}
In Eqs.~(\ref{djs}) the following notations are introduced. $R_0$ is
the zero-field bulk resistivity in neutral graphene
\cite{hydro0,mus}
\begin{equation}
\label{r0}
R_0 =
\frac{\pi}{2e^2T\ln2}\left(\frac{1}{\tau_{11}}+\frac{1}{\tau_{\rm
    dis}}\right) \underset{\tau_{\rm
    dis}\rightarrow\infty}{\longrightarrow} \frac{1}{\sigma_Q},
\end{equation}
where $\tau_{11}\propto\alpha_g^{-2}T^{-1}$ describes the appropriate
electron-electron collision integral. $\tilde{R}_B$ denotes
\cite{megt2,van1}
\begin{eqnarray}
\label{trb}
\tilde{R}_B = \frac{\pi}{2e^2T\ln2}\frac{\omega_B^2}{\tau_{\rm dis}^{-1}\!+\!\delta_I^{-1}\tau_{22}^{-1}},
\end{eqnarray}
where $\tau_{22}\propto\alpha_g^{-2}T^{-1}$ describes a component of
the collision integral that is qualitatively similar, but
quantitatively distinct from $\tau_{11}$ and $\delta_I\approx 0.28$.
Another numerical factor in Eqs.~(\ref{djs}) is
${\alpha_1\approx2.08}$ and ${\omega_B=eBv_g^2/(2cT\ln2)}$ is the
generalized cyclotron frequency at $\mu=0$.

The shear viscosity at charge neutrality and in the absence of
magnetic field was evaluated in Refs.~\cite{msf,me1,me2} and has the form
\begin{equation}
\label{eta0}
\eta(\mu=0, B=0) = {\cal B} \frac{T^2}{\alpha_g^2 v_g^2},
\qquad
{\cal B}\approx 0.45.
\end{equation}
Within the renormalization group (RG) approach, $\alpha_g$ is a
running coupling constant
\cite{shsch,msf,paco99,paconpb,julia}. However, the product
$\alpha_gv_g$ remains constant along the RG flow
\cite{msf,kash}. Hence Eq.~(\ref{eta0}) gives the correct form of
shear viscosity in neutral graphene \cite{shsch}. Within the kinetic
theory approach, the coefficient ${\cal B}$ can be expressed in terms
of time scales characterizing the collision integral
\cite{me1,me2}. At neutrality these time scales are qualitatively
similar to, but quantitatively distinct from $\tau_{11}$ and
$\tau_{22}$. The similarity follows from the fact that in general all
time scales are functions of the chemical potential and temperature
\cite{me1,me3,drag}. At neutrality $\mu=0$ and hence all time scales
are inverse proportional to temperature.

As a function of the magnetic field, the viscosity coefficient in
neutral graphene exhibit a weak decay until
eventually saturating in classically strong fields \cite{me2}
\begin{equation}
\label{etaB}
\eta(\mu=0, B) = \frac{{\cal B}\!+\!{\cal B}_1\gamma_B^2}{1\!+\!{\cal B}_2\gamma_B^2} 
\frac{T^2}{\alpha_g^2 v_g^2},
\qquad
\gamma_B=\frac{|e|v_g^2B}{\alpha_g^2cT^2},
\end{equation}
where
\[
{\cal B}_1\approx 0.0037, 
\qquad
{\cal B}_2\approx 0.0274.
\]
This behavior should be contrasted with the more conventional
Lorentzian decay of field-dependent shear viscosity in doped graphene
\cite{me2,geim4,ale,moo,stein}. However, in weak fields where most
present-day experiments are performed this distinction is
negligible. Moreover, due to the smallness of the coefficient ${\cal
  B}_1$ and ${\cal B}_2$ we disregard the field dependence of $\eta$
in what follows.

Under the assumptions of the hydrodynamic regime, disorder scattering
is characterized by the large mean free time,
$\tau_{\rm{dis}}\gg\tau_{11},\tau_{22}$, yielding a negligible
contribution to Eqs.~(\ref{r0}) and (\ref{trb}). Equation (\ref{r0})
describes the uniform bulk current (at $\bs{B}=0$) and is independent
of viscosity (i.e., in a channel \cite{hydro1,luc,mr1,megt2}). In
contrast, in the Corbino geometry the current flow is necessarily
inhomogeneous and hence viscous dissipation must be taken into
account.

\section{Boundary conditions}

Differential equations (\ref{hydrolin1}) should be supplemented by
boundary conditions, which should be expressed in terms of the
hydrodynamic velocity and macroscopic currents. The statement of the
boundary conditions does not imply the validity of the hydrodynamic
approximation at the sample edges and generally have to be derived
from the underlying microscopic theory. However some of the boundary
conditions can be derived based on the conservation laws alone. In the
circular Corbino geometry conservation laws can be used to establish
boundary conditions for radial components of the currents \cite{van1}.

\subsection{Radial components of macroscopic currents}

A typical experimental setup involves a graphene sample (in our case,
at charge neutrality) in the shape of an annulus placed between the
inner (a disk of radius $r_1$) and outer (a ring with the inner radius
$r_2$) metallic contacts (leads). The electric current $I$ is injected
into the center of the inner lead preserving the rotational invariance
(e.g., through a thin vertical wire attached to the center point) and
spreads towards the outer lead, which for concreteness we assume to be
grounded. The overall voltage drop $U$ is measured between two points
in the two leads (at the radii ${r_{\rm in}<r_1}$ and
${r_{\rm{out}}>r_2}$) yielding the device resistance, ${R=U/I}$. The
only boundaries in the system are between the sample and the external
leads.

For simplicity, we assume both leads to be of the same material with a
single-band electronic system, e.g., highly doped graphene with the
same doping level. In that case, all macroscopic currents in the leads
are proportional to the drift velocity and hence are determined by the
injected current. In the stationary case, the continuity equation
(\ref{cen2}) determines the radial component of the electric current
density. In the inner lead this yields ${j^{\rm{in}}_r=I/(2\pi{e}r)}$,
defining the radial component of the drift velocity,
${u_r^{\rm{in}}=j^{\rm{in}}_r/n_L}$ ($n_L$ is the carrier density in
the inner lead). Assuming charge conservation is not violated at the
interface, we find the boundary condition between the inner lead and
the sample
\begin{subequations}
\label{bcs}
\begin{equation}
\label{bcjrin}
j_r(r_1\!-\!\epsilon) = n_L u_r(r_1\!-\!\epsilon) = \delta j_r (r_1\!+\!\epsilon),
\end{equation}
where $\epsilon>0$ is infinitesimal and we took into account that in
neutral ($n=0$) graphene $\bs{j}=\delta\bs{j}$.

The second hydrodynamic equation, Eq.~(\ref{ceni2}), is the continuity
equation for the imbalance density. Although the total quasiparticle
number is not conserved, integrating this equation over an
infinitesimally thin region encompassing the boundary yields a similar
boundary condition for the imbalance current assuming that the
relaxation rate due to quasiparticle recombination is not singular at
the boundary
\begin{equation}
\label{bcjIrin}
j_{I,r}(r_1\!-\!\epsilon) = n_L u_r(r_1\!-\!\epsilon) 
= 
n_I u_r(r_1+\epsilon) + \delta j_{I,r} (r_1+\epsilon).
\end{equation}
Here we took into account the fact that in a single-band system
$\bs{j}_I$ is identical with $\bs{j}$.

Finally, Eq.~(\ref{tteqlin1}) is the linearized continuity equation
for the entropy density (here we follow the standard practice
\cite{dau6} of replacing the continuity equation for the energy
density by the entropy flow equation, also known as the thermal
transport equation). Again, assuming the energy relaxation rate is not
singular at the interface (i.e., the current flow is not accompanied
by energy or excess heat accumulation at the boundary between the
sample and the contact) we integrate Eq.~(\ref{tteqlin1}) over an
infinitesimally thin region encompassing the boundary and arrive at
the boundary condition for the entropy current
\begin{equation}
\label{bcjenrin}
s^{\rm{in}} u_r(r_1\!-\!\epsilon) = s u_r(r_1+\epsilon),
\end{equation}
\end{subequations}
where $s$ and $s^{\rm{in}}$ are the entropy densities in the sample
and inner lead, respectively.

\subsection{Tangential flows in external magnetic field}
\label{bctheta}

The above boundary conditions (and the corresponding conditions on the
outer lead) are sufficient to solve the hydrodynamic equations in the
absence of magnetic field where all currents are radial \cite{van1}.
An external magnetic field induces the tangential components of the
currents due to the classical Hall effect. The continuity equations do
not determine the tangential components and hence the boundary
conditions have to be derived from a microscopic theory. Generally
speaking, the boundary conditions depend on the presence of tangential
forces at the interface, usually associated with edge
roughness. Typically \cite{dau6,me0,luc,rev,fal19}, one considers the
two limiting cases of either the ``no-slip'' or ``no-stress'' boundary
conditions corresponding to either the presence or the absence of the
drag-like friction across the interface.

For contact interfaces in the Corbino geometry, the boundary
conditions corresponding to the above limiting cases differ from the
well-known expression of conventional hydrodynamics. The no-slip
boundary condition now means that the tangential component of the
hydrodynamic velocity is continuous across the interface (written as
above for the inner interface)
\begin{subequations}
\label{bcut}
\begin{equation}
\label{noslip}
u_{L\vartheta}(r_1-\epsilon)=u_\vartheta(r_1+\epsilon),
\end{equation}
in contrast to the common condition of vanishing velocity at the
channel boundary (the two are consistent, since in the latter case
there is no flow beyond the edge). 

The no-stress boundary condition means the absence of any forces along
the interface in which case the tangential component of the stress
tensor $\Pi^{ij}$ is continuous. In polar coordinates appropriate for
the Corbino geometry one finds
\begin{equation}
\label{nostress}
\Pi_{L,E}^{\vartheta r}(r_1-\epsilon)=\Pi_E^{\vartheta r}(r_1+\epsilon),
\end{equation}
\end{subequations}
The no-stress boundary condition is easy to derive starting from the
kinetic equation. Multiplying the kinetic equation by the momentum and
summing over all quasiparticle states, one finds an equation featuring
the gradient of the stress tensor \cite{me1} as well as macroscopic
forces in the system. Now the boundary condition can be obtained by
integrating that equation over the small volume around the
interface. Unless there is a force localized at the interface (with a
$\delta$-function-like coordinate dependence on the hydrodynamic
scale), this procedure would yield Eq.~(\ref{nostress}). Usually, the
interfaces are microscopically rough with the roughness providing such
a force. As a result, the no-slip boundary condition is more commonly
used. In neutral graphene, however, the quasiparticle wavelength
typically exceeds any length scale associated with edge roughness
leading to specular scattering \cite{megt2} and Eq.~(\ref{nostress}).

In the case of the hard wall edges, the boundary conditions were
previously studied theoretically in Ref.~\cite{ks19} and confirmed
experimentally in Ref.~\cite{imm} where a nonzero slip length was
proposed indicating a more general Maxwell's boundary
condition. However, the specific choice of the boudnary conditions
does not lead to qualitatively different results \cite{fal19}. Here we
follow the hydrodynamic tradition and consider both the no-slip and
no-stress boundary conditions.

\subsection{Interface-induced dissipation and jumps of the electric potential}

The hydrodynamic theory discussed so far completely describes the
energy flow in neutral graphene. In order to establish the device
resistance $R$ we have to find the behavior of the electrochemical
potential at the interfaces.

The standard description of interfaces between metals or
semiconductors in terms of the contact resistance \cite{mard} can be
carried over to neutral graphene \cite{alf}. In graphene, the contact
resistance was recently measured in Ref.~\cite{imm} (see also
Refs.~\cite{sulp22,gall21,hak}). In the diffusive (or Ohmic) case, the
contact resistance leads to a voltage drop that is small compared to that
in the bulk of the sample and can be neglected. In contrast, in the
ballistic case with almost no voltage drop in the bulk, most energy is
dissipated at the contacts. Both scenarios neglect interactions.

\begin{figure}[t]
\centering
\includegraphics[width=\columnwidth]{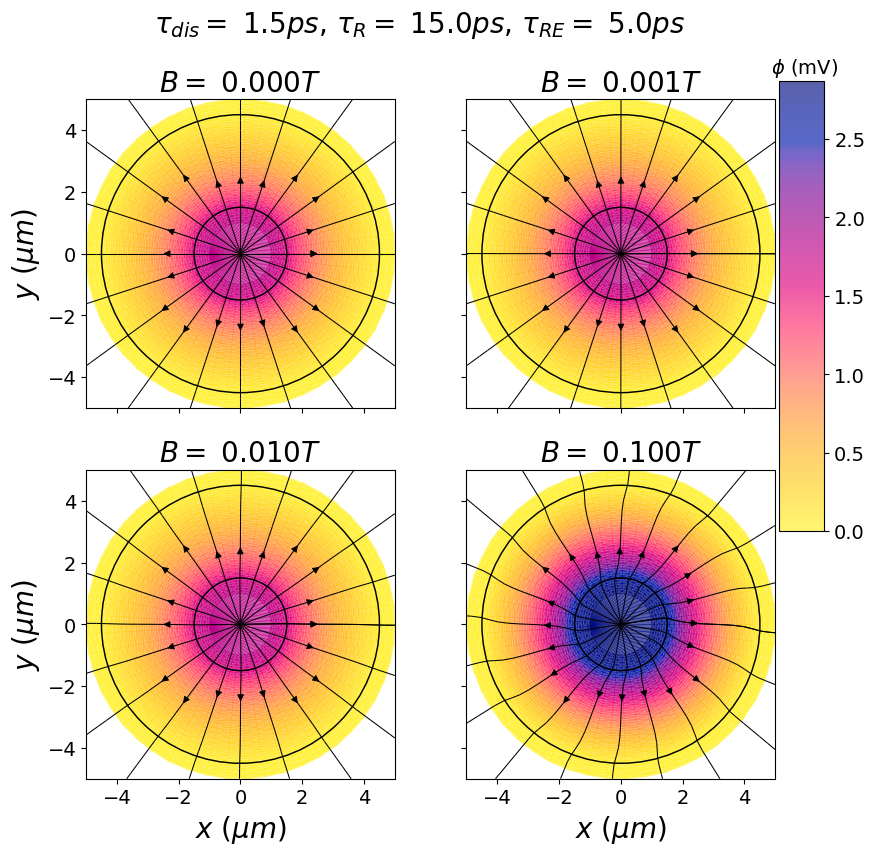}
\caption{Electric current density $\bs{j}$ and potential $\varphi$ in
  the device obtained by solving the hydrodynamic equations at
  relatively high temperatures where energy relaxation is dominated by
  supercollisions. Arrows indicate $\bs{j}$ and the color map shows
  $\varphi$. The outer lead is chosen to be grounded. The four panels
  correspond to the indicated values of magnetic field. For the values
  of other parameters, see Fig~\ref{fig:ut}.}
\label{fig:jf}
\end{figure}

In the diffusive regime interactions give rise to perturbative
corrections to the bulk resistivity \cite{aar,zna} and the contact
resistance can still be neglected. In ballistic samples
electron-electron interaction may lead to the ``Knudsen-Poiseuille''
crossover \cite{gurzhi} and drive the system to the hydrodynamic
regime. In this case the Ohmic resistivity of the electronic fluid may
remain small, but there exist other channels for dissipation due to
viscosity \cite{fal19} and energy relaxation processes
\cite{meig1}. In neutral graphene the effect is subtle \cite{van1},
since the electric current is decoupled from the hydrodynamic energy
flow. However, both are induced by the current source that provides
the energy dissipated through all the above channels. The energy
dissipated in the system corresponds to the overall voltage drop. In
the bulk of the sample the voltage drop is Ohmic as determined by
Eq.~(\ref{lj}), while the additional contribution takes the form of a
potential jump at the interface between the sample and leads. At the
same time, an excess electric field is induced in a thin Knudsen layer
around the interface \cite{fal19}.

The magnitude of the jump in $\phi$ can be established by considering
the flow of energy through the interface as suggested in
Ref.~\cite{fal19} and detailed in neutral graphene at $\bs{B}=0$ in
Ref.~\cite{van1}. Consider the kinetic energy defined by integrating
the energy density ${n_E(\bs{u})\!-\!n_E(0)}$ over the volume
\begin{equation}
\label{ke}
{\cal E} = \int dV \left[n_E(\bs{u})\!-\!n_E(0)\right]
\approx \int dV \frac{6P}{v_g^2} u^2,
\end{equation}
which we have expanded to the leading order in $\bs{u}$ (and hence in
$I$). In the stationary state, dissipation is balanced by the work
done by the source, such that the time derivative of the kinetic
energy vanishes, ${{\cal A}=\dot{\cal E}=0}$. Using the equations of
motion and continuity equations to find time derivatives, one may
split ${\cal A}$ into the ``bulk'' and ``boundary'' contributions,
${\cal A}={\cal A}_{\rm bulk}+{\cal A}_{\rm edge}$. The former may be
interpreted as the bulk dissipation, while ${\cal A}_{\rm edge}$ must
include the energy brought in (carried away) through the boundary by
the incoming (outgoing) flow. The boundary condition is then found
under the assumption that energy is not accumulated at the interface.

Assuming the leads' material is highly doped graphene, the equation of
motion is the usual Ohm's law where we may combine the diffusion term
\cite{df2} with a contribution of viscosity $\eta_L$ due to disorder
\cite{bur19} into the gradient of the stress-energy tensor
\cite{hydro0} and hence
\begin{eqnarray*}
&&
\!\!\frac{3P_L}{v_g^2}\bs{u}_L\partial_t\bs{u}_L =
\nonumber\\
&&
\nonumber\\
&&
\quad
=
u^i_L\left(
-\frac{3P_L}{v_g^2}\frac{u^i_L}{\tau_L}\!-\!\nabla^j\Pi^{ij}_{L,E}
\!+\!n_L e E^i\!+\!\frac{e}{c}\epsilon^{ijk}j^jB^k
\right)
\nonumber\\
&&
\nonumber\\
&&
\quad
=-\frac{3P_L}{v_g^2}\frac{u_L^2}{\tau_L}
\!+\!\frac{\partial u_{L,i}}{\partial x_j}\Pi_{L,E}^{ij}
\!+\!\frac{e}{c}\bs{u}_L\!\cdot\!(\bs{j}\!\times\!\bs{B})
\!+\!e\varphi\bs{\nabla}\!\cdot\!\bs{j}
\nonumber\\
&&
\nonumber\\
&&
\qquad\qquad
-\nabla^i\left(u^j_L\Pi_{L,E}^{ij}+ej^i\varphi\right).
\end{eqnarray*}
The last term in this expression determines the boundary
contribution. Given that the Lorentz force does not explicitly
contribute, the only difference from the expression derived in
Ref.~\cite{van1} at $\bs{B}=0$ is the nonzero tangential components
of the hydrodynamic velocity and the stress tensor (vanishing in the
absence of magnetic field). In neutral graphene, we obtain a similar
expression from the Navier-Stokes equation, while the Joule heat is
determined by $\delta\bs{j}$. Equating the two contributions we find
the jump of the potential in the form 
\begin{eqnarray}
\label{bcphi}
&&
\varphi(r_1-\varepsilon)-\varphi(r_1+\varepsilon)
= I R_c +
\\
&&
\nonumber\\
&&
\qquad\qquad
+\frac{2\pi r_1}{I}\left[\left(u_r\Pi_{E}^{rr}+u_\vartheta\Pi_{E}^{\vartheta r}\right)\Big|_{r_1+\varepsilon}
\right.
\nonumber\\
&&
\nonumber\\
&&
\qquad\qquad\qquad\qquad
\left.
-\left(u_r\Pi_{L,E}^{rr}+u_\vartheta\Pi_{L,E}^{\vartheta r}\right)\Big|_{r_1-\varepsilon} \right]\!,
\nonumber
\end{eqnarray}
where $R_c$ is the usual contact resistance \cite{alf}. A similar
condition holds at the boundary with the outer lead.

\section{Hydrodynamic flows in the Corbino geometry}

In polar coordinates and taking into account radial symmetry, the
hydrodynamic equations (\ref{hydrolin1}) and (\ref{djs}) form two
disjoint sets of differential equations. The first one determines the
tangential component of the velocity $u_\vartheta$:
\begin{subequations}
\label{eqs1}
\begin{equation}
\label{eq:hyd1}
\frac{1}{r}\frac{\partial(r\delta j_r)}{\partial r} = 0,
\end{equation}
\begin{equation}
\label{eq:hyd4}
\eta \partial_r\left(\frac{1}{r}\frac{\partial(ru_\vartheta)}{\partial r}\right)
-
\frac{eB}{c}\delta j_r
-
\frac{3Pu_\vartheta}{v_g^2\tau_{{\rm dis}}} = 0,
\end{equation}
\begin{equation}
\label{eq:dis1}
\delta j_r = 
\frac{1}
{e^2\tilde{R}}
\left[ eE_r(r)
+
\omega_B
\frac{2T\ln2}{v_g^2}u_\vartheta\right]\!,
\end{equation}
\begin{equation}
\label{eq:dis4}
\delta j_{I\vartheta} = 
-\frac{\alpha_1\delta_I\omega_B}{\tau_{\rm dis}^{-1}\!+\!\delta_I^{-1}\tau_{22}^{-1}}\delta j_r,
\end{equation}
\end{subequations}
while the second one involves the radial component $u_r$:
\begin{subequations}
\label{eqs2}
\begin{equation}
\label{eq:hyd2}
\frac{n_{I}}{r}\frac{\partial(ru_r)}{\partial r} 
+ \frac{1}{r}\frac{\partial(r \delta j_{Ir})}{\partial r}
= -\frac{12\ln2}{\pi^2}\frac{n_{I}\mu_I(r)}{T\tau_R},
\end{equation}
\begin{equation}
\label{eq:hyd3}
\frac{\partial \delta P}{\partial r}
=
\eta \partial_r\left(\frac{1}{r}\frac{\partial(ru_r)}{\partial r}\right)
+
\frac{eB}{c}\delta j_\vartheta
-
\frac{3Pu_r}{v_g^2\tau_{{\rm dis}}},
\end{equation}
\begin{equation}
\label{eq:hyd5}
\frac{3P}{r}\frac{\partial(r u_r)}{\partial r} = - \frac{2\delta{P}(r)}{\tau_{RE}}.
\end{equation}
\begin{equation}
\label{eq:dis2}
\delta j_\vartheta = 
\frac{\omega_B}
{e^2\tilde{R}}
\left(
\frac{\alpha_1\delta_I}{\tau_{\rm dis}^{-1}\!+\!\delta_I^{-1}\tau_{22}^{-1}}\frac{\partial \mu_I}{\partial r}
-
\frac{2T\ln2}{v_g^2}u_r\right)\!,
\end{equation}
\begin{equation}
\label{eq:dis3}
\delta j_{Ir} = 
-\frac{2\delta_IT\ln2}{\tau_{\rm dis}^{-1}\!+\!\delta_I^{-1}\tau_{22}^{-1}}
\left[
\frac{R_0}{\pi\tilde{R}}\frac{\partial \mu_I}{\partial r}
\!+\!
\frac{\alpha_1\omega_B^2}{e^2\tilde{R}}\frac{u_r}{v_g^2}
\right]\!.
\end{equation}
\end{subequations}

\subsection{Tangential component of the velocity and bulk voltage drop}
\label{sec_rb}

The bulk magnetoresistance can be found by solving Eqs.~(\ref{eqs1})
with the appropriate boundary conditions. Combining
Eqs.~(\ref{eq:hyd1}) and (\ref{eq:hyd4}) we find an inhomogeneous
Bessel equation for the tangential component of the velocity
$u_\vartheta$ with the characteristic length scale being the Gurzhi
length $\ell_G^2=\eta v_g^2\tau_{\rm dis}/(3P)$. The boundary
condition for $u_\vartheta$ is determined by microscopic details of
viscous drag at the interface and hence is not universal. Here we
follow the hydrodynamic tradition and consider both the no-slip and
the no-stress boundary conditions, see Sec.~\ref{bctheta}. Moreover, one
can distinguish two different setups where the external magnetic field
is applied either to the sample only or to the whole device including
the leads. In all these cases we can find an analytic expression for
$u_\vartheta$, which can be substituted into of Eq.~(\ref{eq:dis1}) to
find the electric field in the sample, $E_r$ (the radial component of
the current is determined by the continuity equation
alone). Similarly, Eq.~(\ref{eq:dis4}) determines $\delta
j_{I\vartheta}$.  Using the obtained electric field we can determine
the voltage drop through the bulk of the sample as
\begin{equation}
U=\int\limits_{r_1}^{r_2} E_r dr 
= 
\int\limits_{r_1}^{r_2}dr\left( \frac{\tilde{R}I}{2\pi r}-\frac{B}{c}u_\vartheta \right).
\end{equation}

For the no-slip boundary condition for $u_\vartheta$ and allowing the
external magnetic field to penetrate the leads, the tangential
component of the velocity is given by
\begin{eqnarray}
&&
u_\vartheta=-\frac{B I \ell_G^2}{2 \pi  c\eta r} +
\frac{BI\left(\eta\ell_L^2-\eta_L\ell_G^2\right)}{2\pi c\eta \eta_Lr_1r_2}\times
\\
&&
\nonumber\\
&&
\quad
\times\left[K_1\!\left(\frac{r}{\ell_G}\right)
\frac{r_1I_1\!\left(\frac{r_1}{\ell_G}\right)\!-\!r_2I_1\!\left(\frac{r_2}{\ell_G}\right)}
{K_1\!\left(\frac{r_1}{\ell_G}\right)I_1\!\left(\frac{r_2}{\ell_G}\right)
\!-\!I_1\!\left(\frac{r_1}{\ell_G}\right) K_1\!\left(\frac{r_2}{\ell_G}\right)}
\right.
\nonumber\\
&&
\nonumber\\
&&
\qquad
+
\left.
I_1\!\left(\frac{r}{\ell_G}\right)
\frac{r_2 K_1\!\left(\frac{r_2}{\ell_G}\right)\!-\!r_1 K_1\!\left(\frac{r_1}{\ell_G}\right)}
{K_1\!\left(\frac{r_1}{\ell_G}\right) I_1\!\left(\frac{r_2}{\ell_G}\right)
\!-\!I_1\!\left(\frac{r_1}{\ell_G}\right) K_1\!\left(\frac{r_2}{\ell_G}\right)}
\right]\!,
\nonumber
\end{eqnarray}
where $\eta_L$ is the disorder-induced viscosity \cite{bur19} and
$\ell_L^2=v_g^2\eta_L\tau_L/(2P_L)$ is the Gurzhi length
in the leads.

In the limit $\ell_G\gg r_1,r_2$ (i.e., ``clean system'' with long
mean free time $\tau_{\rm dis}\rightarrow\infty$) this simplifies to
($p=r_2/r_1$)
\begin{eqnarray}
&&
\!\!\!\!
u_\vartheta \approx -\frac{BI\ell_L^2}{4\pi cr\eta_L} 
\left[2+\left(\frac{1}{\ell_G^2}\!-\!\frac{\eta_L}{\eta\ell_L^2}\right)
\right.\times
\\
&&
\nonumber\\
&&
\qquad\qquad\qquad\quad
\times
\left.
\frac{r^2\ln\left(\frac{r}{r_1}\right)
\!+\!r^2p^2\ln\left(\frac{r_2}{r}\right)\!-\!r_2^2\ln p}
{1\!-\!p^2}\right]\!.
\nonumber
\end{eqnarray}
The corresponding voltage drop remains finite 
\begin{subequations}
\label{ub1}
\begin{eqnarray}
&& 
U\approx  \left(1\!-\!\frac{\eta\ell_L^2}{\eta_L\ell_G^2}\right) \frac{B^2Ir_2^2}{4\pi c^2\eta} 
\frac{ (p^2\!-\!1)^2\!-\!4p^2 \ln^2p}{4p^2(p^2\!-\!1)}
\\
&&
\nonumber\\
&&
\qquad\qquad\qquad\qquad\qquad\qquad
+\frac{I\ln p}{2\pi} \left(\frac{B^2}{c^2}\frac{v_g^2\tau_L}{3P_L}+\tilde{R}\right)\!,
\nonumber
\end{eqnarray}
yielding the field-dependent bulk resistance ($R=U/I$)
\begin{eqnarray}
&& 
R(B)\approx  \frac{\ln p}{2\pi} R_0 + \frac{B^2 r_2^2}{4\pi c^2\eta}
\frac{(p^2\!-\!1)^2\!-\!4p^2 \ln^2p}{4p^2(p^2\!-\!1)}
\\
&&
\nonumber\\
&&
\qquad\qquad
+
\frac{B^2v_g^4\ln p}{2c^2T^3}\left[
\frac{\alpha_1^2\delta_I}{8\ln^32}\frac{1}{\tau_{\rm dis}^{-1}\!+\!\delta_I^{-1}\tau_{22}^{-1}}
+
\frac{T^3}{\mu^3}\tau_L
\right]\!,
\nonumber
\end{eqnarray}
\end{subequations}
assuming
${\eta\ell_L^2/(\eta_L\ell_G^2)=3P\tau_L/(2P_L\tau_{\rm{dis}})\ll1}$
with $P_L=\mu_L^3/(3\pi v_g^2)$. The two field-dependent terms differ
in their dependence on temperature, sample size, and coupling constant
\cite{geim5} opening a possibility to separate the two contributions
from the experimental data and thus to measure the viscosity
coefficient.

If the magnetic field is applied to the sample only (and not to the
leads) $u_\vartheta$ vanishes in the leads and hence the terms with
$\ell_L$ do not appear in the voltage drop (\ref{ub1}). In that case,
the field-dependent contribution to $U$ does not contain
$\tau_{\rm{dis}}$ in contrast to the known result in the strip
geometry \cite{hydro0,mus}.

A similar result can be obtained in the case of no-stress boundary
conditions, where the tangential component of the velocity
$u_\vartheta$ is still expressed in terms of the Bessel functions. In
the clean limit ($\ell_G\gg r_1,r_2$) the voltage drop also remains
finite
\begin{eqnarray}
\label{ub2}
&& 
U\approx  
\frac{I}{2\pi} \left(\tilde{R}+\frac{B^2\ell_L^2}{c^2\eta}-\frac{B\eta_H}{ec\eta n_L}\right) \ln p
\\
&&
\nonumber\\
&&
\qquad\qquad
+\frac{r_2^2B^2I}{4\pi c^2\eta} 
\left[\frac{ \left(p^2\!-\!1\right) \left(p^4\!+\!10p^2\!+\!1\right)}{12p^2\left(p^2\!+\!1\right)^2}
-\frac{\ln p}{1\!+\!p^2}\right]
\nonumber\\
&&
\nonumber\\
&&
\qquad\qquad
+\frac{I}{2\pi} 
\left[\frac{B^2}{c^2}\frac{\left(\ell_G^2\!-\!\ell_L^2\right)}{\eta}+\frac{B\eta_H}{ec\eta n_L}\right]
\frac{p^2\!-\!1}{p^2\!+\!1},
\nonumber
\end{eqnarray}
where $\eta_H$ is the Hall viscosity in the leads, which vanishes if
the magnetic field is not allowed in the leads. In that case, the last
term in the voltage drop (\ref{ub2}) is proportional to
$\tau_{\rm{dis}}$ and independent of viscosity. The second term in
Eq.~(\ref{ub2}) remains similar to Eq.~(\ref{ub1}) and is inverse
proportional to $\eta$. This term's dependence on the ratio $p$ is
distinct from both Eq.~(\ref{ub1}) and the third term in
Eq.~(\ref{ub2}) and could be extracted by analyzing the data in a set
of Corbino disks with different $p$.

\begin{figure}[t]
\centering
\includegraphics[width=\columnwidth]{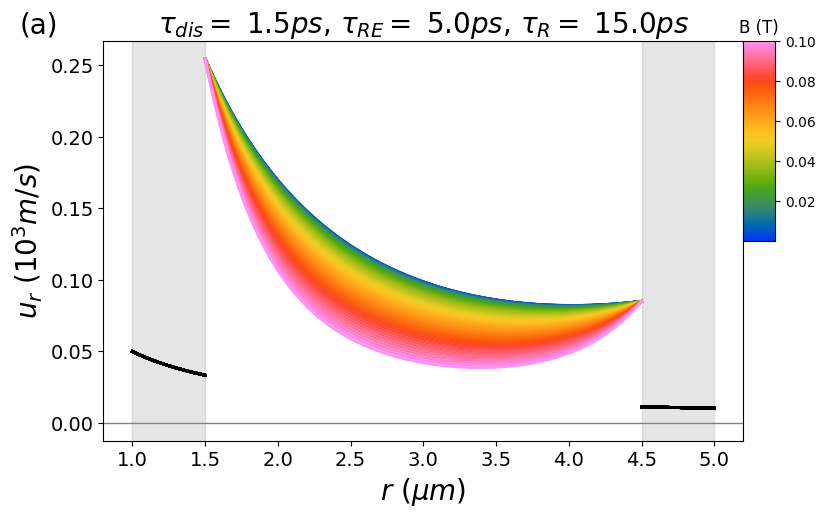}
\includegraphics[width=\columnwidth]{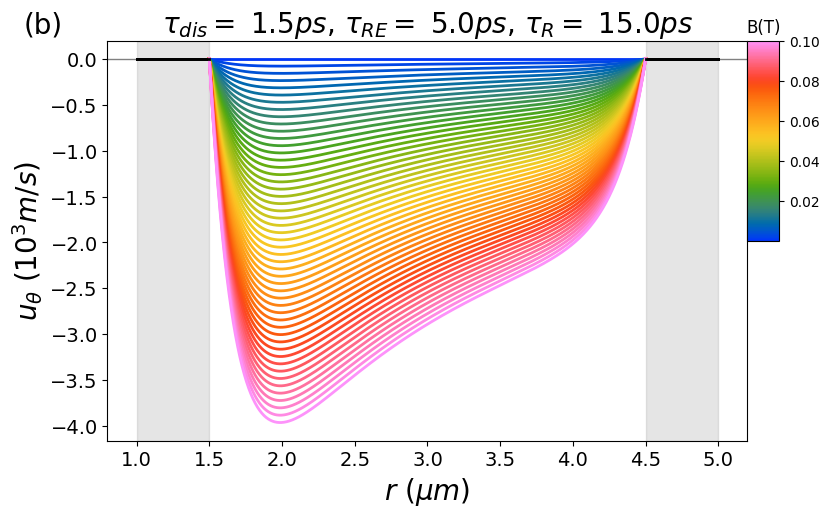}
\caption{Radial (top panel) and tangential (bottom panel) components
  of the hydrodynamic velocity $\bs{u}$ computed within the
  ``supercollisions model'' of energy relaxation. Black lines in the
  shaded regions show the drift velocity in the leads. Color curves
  correspond to different values of the external magnetic field
  according to the shown color coding. The top curve shows values at
  $\bs{B}=0$ and is identical with the results of
  Ref.~\cite{van1}. For the parameter values, see Fig~\ref{fig:ut}.}
\label{fig:us}
\end{figure}

In the opposite limit $\ell_G\ll r_1,r_2$, the leading contribution to
the bulk voltage drop is independent of $\eta$. For no-slip boundary
conditions and in the simplified case where the field is not allowed
to penetrate the leads we find for the field-dependent part of $U$
\begin{equation}
\label{ub3}
R(B)\!-\!R(0) \approx \frac{B^2v_g^2\tau_\text{dis}\ln p}{6\pi c^2P}
+\frac{\ln p}{2\pi} \delta_I\alpha_1^2 \tilde{R}_B\propto \tau_\text{dis}B^2.
\end{equation}
The voltage drop (\ref{ub3}) is proportional to $\tau_{\rm{dis}}$
similarly to the result in the strip geometry (see
Refs.~\cite{hydro0,mus}). Of course, in the limit $\ell_G\ll
r_1,r_2$ the mean free time $\tau_{\rm{dis}}$ cannot be arbitrarily
large, hence the voltage drop (\ref{ub3}) does not diverge. In the
limit $\tau_{\rm{dis}}\rightarrow\infty$ the voltage drop crosses over
to the above ``clean'' limit and is given by Eq.~(\ref{ub1}). However,
the limiting expression (\ref{ub3}) is independent of viscosity, and
hence qualitatively similar to the usual result.

To summarize the results of this section, we have shown that bulk
magnetoresistance in neutral graphene in the Corbino geometry exhibits
a crossover between the ``clean'' limit of the large (compared to the
disk radius) Gurzhi length to the limit of small Gurzhi length. In the
former case, the field-dependent part of the bulk voltage drop is
determined by viscosity, while in the latter limit it is proportional
to the disorder mean free time similarly to the known result in the
strip geometry.

\subsection{Radial component of the velocity and the device resistance}

The five equations (\ref{eqs2}) can be reduced to two coupled
differential equations (for similar calculations in the strip geometry
see Refs.~\cite{hydro0,mr2,mr3,megt2}). To simplify the arguments, we
introduce the following notations
\begin{equation}
\label{ns1}
q = n_I u_r,
\quad
p = \delta j_{I,r},
\quad
x = \frac{2n_I}{3P} \delta P,
\quad
y = \frac{12\ln 2}{\pi^2} \frac{n_I}{T} \mu_I.
\end{equation}
In terms of the new variables, Eqs.~(\ref{eq:hyd2}) and
(\ref{eq:hyd5}) can be written as
\begin{subequations}
\label{2eq1}
\begin{equation}
\frac{1}{r}\frac{\partial (rq)}{\partial r}
+
\frac{1}{r}\frac{\partial (rp)}{\partial r}
=
-\frac{y}{\tau_R},
\end{equation}
\begin{equation}
\frac{1}{r}\frac{\partial (rq)}{\partial r} = -\frac{x}{\tau_{RE}}.
\end{equation}
\end{subequations}
Equation (\ref{eq:dis3}) can be rewritten as
\begin{subequations}
\label{2eq2}
\begin{equation}
\frac{\partial y}{\partial r} = - \frac{6}{\pi} \frac{\tilde R n_I}{R_0 T^2 \tilde\tau} \, p
-
\frac{12\ln 2}{\pi} \frac{\alpha_1\omega_B^2}{e^2v_g^2 R_0 T} \, q,
\end{equation}
where ${\tilde\tau=\delta_I/(\tau_{\rm
    dis}^{-1}\!+\!\delta_I^{-1}\tau_{22}^{-1})}$. Finally,
Eqs.~(\ref{eq:hyd3}) and (\ref{eq:dis2}) can be combined into
\begin{eqnarray}
&&
\frac{\partial x}{\partial r} = \frac{2\eta}{3P} 
\frac{\partial}{\partial r}\frac{1}{r}\frac{\partial (rq)}{\partial r}
-
\frac{2}{v_g^2}\left[\tau_{\rm dis}^{-1}+\frac{\omega_B^2}{e^2\tilde R}\frac{4T^2\ln^22}{3Pv_g^2}\right]q
\nonumber\\
&&
\nonumber\\
&&
\qquad\qquad\qquad\qquad\qquad
+
\alpha_1\tilde\tau\frac{\pi^2T^2}{9Pv_g^2}\frac{\omega_B^2}{e^2\tilde R}\frac{\partial y}{\partial r}.
\end{eqnarray}
\end{subequations}
Introducing the differential operator
\begin{equation}
\hat{\mathbb{D}}q = \frac{\partial}{\partial r}\frac{1}{r}\frac{\partial (rq)}{\partial r},
\end{equation}
we rewrite Eqs.~(\ref{2eq1}) in the matrix form
\begin{subequations}
\label{2eq}
\begin{equation}
\label{ts}
\hat{\mathbb{D}}
\begin{pmatrix}
q \cr
p
\end{pmatrix}
=
\widehat{T}_S 
\begin{pmatrix}
\partial x/\partial r \cr
\partial y/\partial r
\end{pmatrix}
,
\qquad
\widehat{T}_S
=
\begin{pmatrix}
\frac{1}{\tau_\text{RE}}  & 0 \cr
-\frac{1}{\tau_\text{RE}} & \frac{1}{\tau_R}
\end{pmatrix}.
\end{equation}
Similarly, Eqs.~(\ref{2eq2}) can be written in the matrix form
\begin{equation}
\begin{pmatrix}
\partial x/\partial r \cr
\partial y/\partial r
\end{pmatrix}
= - \widehat{M} 
\begin{pmatrix}
q \cr
p
\end{pmatrix}
+
\widehat{V}\hat{\mathbb{D}}
\begin{pmatrix}
q \cr
p
\end{pmatrix},
\end{equation}
\end{subequations}
where
\[
\widehat{V} = 
\begin{pmatrix}
\frac{2 \eta }{3 P} & 0 \cr
 0 & 0 
\end{pmatrix},
\]
and
\[
\widehat{M} =
\begin{pmatrix}
\frac{16\ln^32}{3\pi}\frac{\delta_I\tilde{R}_BT^3}{v_g^4PR_0\tilde\tau}+\frac{2}{v_g^2\tau_\text{dis}} 
& \frac{4\ln 2}{3} \frac{\alpha_1\delta_In_I\tilde{R}_BT}{v_g^2PR_0\tilde\tau} \cr
 \frac{24\ln^22}{\pi^2}\frac{\alpha_1\delta_I\tilde{R}_B}{v_g^2 R_0 \tilde\tau} 
& \frac{6}{\pi}\frac{n_I\tilde R}{R_0 T^2 \tilde\tau} 
\end{pmatrix}.
\]

\begin{figure}[t]
\centering
\includegraphics[width=\columnwidth]{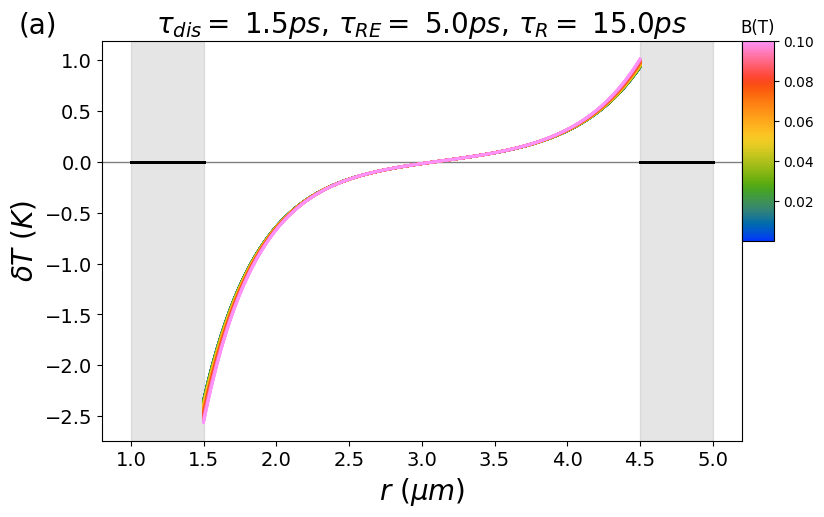}
\includegraphics[width=\columnwidth]{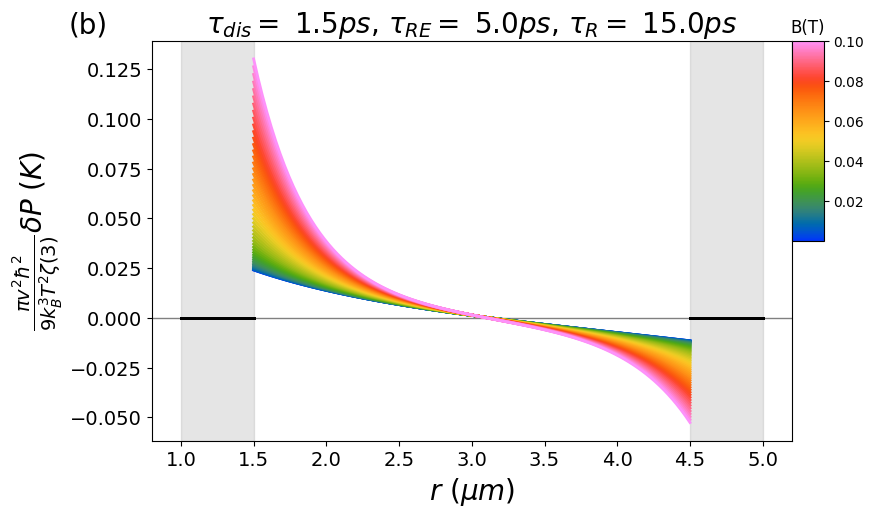}
\caption{Local variations of temperature (top panel) and pressure
  (bottom panel) in the Corbino device computed within the
  ``supercollisions model'' of energy relaxation. Black lines in the
  shaded regions indicate that the leads are at equilibrium.
  Color curves correspond to different values of the external magnetic
  field according to the shown color coding. Zero field values are
  identical with the results of Ref.~\cite{van1}. For the parameter
  values, see Fig~\ref{fig:ut}.}
\label{fig:tps}
\end{figure}

Finally, combining Eqs.~(\ref{2eq}) we find the equation for the
variables $p$ and $q$
\begin{equation}
\label{eq2}
\hat{\mathbb{D}}
\begin{pmatrix}
q \cr
p
\end{pmatrix}
=
\widehat{K}
\begin{pmatrix}
q \cr
p
\end{pmatrix},
\quad
\widehat{K}=\left[1-\widehat{T}_S\widehat{V}\right]^{-1}\widehat{T}_S\widehat{M}.
\end{equation}
The obtained equation should be solved with the boundary conditions
(\ref{bcs}). The solution is straightforward albeit tedious. The
results can be expressed in terms of linear combinations of the Bessel
functions. Thus obtained solutions are not particularly instructive,
hence we present the results of the calculation in graphical form.

The radial component of the hydrodynamic velocity is shown in the top
panel of Fig.~\ref{fig:us}. The drift velocity in the leads shows the
standard Corbino profile, $u_r\propto1/r$. At each interface, $u_r$
exhibits a jump due to the mismatch of the entropy densities in the
sample and leads. For high enough magnetic field, $u_r$ has a sign
change close to the interface. However, the corresponding change of
direction is hardly seen in the overall flow diagram shown in
Fig.~\ref{fig:ut}, since the numerical value of the tangential
component $u_\vartheta$ is much larger (see the bottom panel of
Fig.~\ref{fig:us}).

\begin{figure}[t]
\centering
\includegraphics[width=\columnwidth]{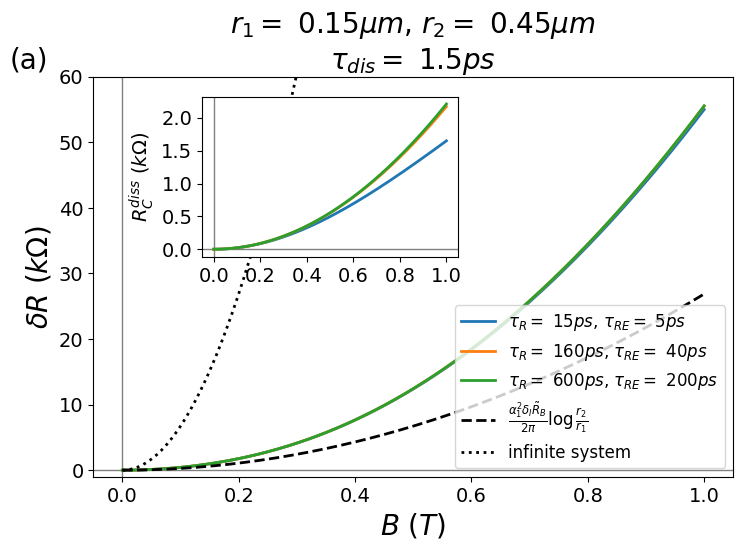}
\includegraphics[width=\columnwidth]{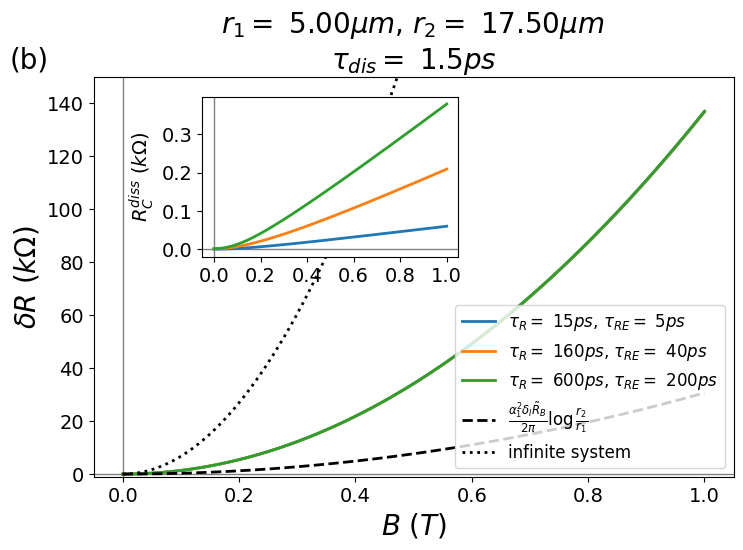}
\caption{Magnetoresistance of a small (top panel) and large (bottom
  panel) Corbino device computed within the ``supercollisions model''
  of energy relaxation. The radii of the Corbino disks are shown above
  the plots. The black dotted line shows the quantity $\tilde R$, which is
  of the same order of magnitude as the magnetoresistance in the
  infinite system \cite{hydro0,megt2}. Color curves correspond to
  three different sets of values of the relaxation times. For other
  parameter values (yielding $\ell_G=0.2\,\mu$m), see
  Fig~\ref{fig:ut}. The insets show the contact resistance due to
  viscous dissipation.}
\label{fig:rs}
\end{figure}

The hydrodynamic velocity determines the energy current in the
system. The nonuniform energy current results in local variations of
the electronic temperature from its equilibrium value (see
Fig.~\ref{fig:tps}). The inhomogeneous temperature profile suggests
that energy relaxation is less effective in strong magnetic fields.
Fig.~\ref{fig:ut} shows the same data as Fig.~\ref{fig:tps} but in the
form of the color map.

Finally we use the boundary conditions (\ref{bcphi}) to find the
interface jumps of the electric potential which allows us to determine
the device resistance. The procedure is the same as in the case of
$\bs{B}=0$ discussed in Ref.~\cite{van1}. The results are shown in
Fig.~\ref{fig:rs}. For small enough samples (see the top panel in
Fig.~\ref{fig:rs}) the device resistance deviates only slightly from
$\tilde R$ which is of the same order of magnitude as the
magnetoresistance in the infinite system \cite{hydro0,megt2}. In large
samples the deviation is more pronounced and depends on the actual
radius of the disk rather than on the ratio $p$ (which is the same in
both plots).

The quantitative results shown in this section were computed for a
particular choice of the relaxation times. These values are largely
phenomenological; however, the magnetoresistance shown in
Fig.~\ref{fig:rs} hardly depends on them, while for larger samples
(the bottom panel) the three curves are indistinguishable. However,
the values of the relaxation times cannot be completely arbitrary. The
point is that the matrix $\widehat{K}$ in Eq.~(\ref{eq2}) is not
guaranteed to have real, positive eigenvalues (although its
determinant is positive). In particular, the recombination time
$\tau_R$ and energy relaxation time $\tau_{RE}$ cannot be very
different. Within the physical model of supercollisions \cite{meig1}
these time-scales are of the same order of magnitude. Quasiparticle
recombination involves supercollision scattering between the bands,
while energy relaxation includes an additional contribution of intraband
scattering. As a result, the energy relaxation time is shorter than
$\tau_R$, but not much shorter since the model does not involve any
additional parameter. For such physical values of the relaxation times
the eigenvalues of the matrix $\widehat{K}$ are real positive and the
resulting magnetoresistance is well accounted for by the curves shown
in Fig.~\ref{fig:rs} where, again, the particular values of $\tau_R$
and $\tau_{RE}$ do not have a strong quantitative impact on the
overall resistance magnitude.

\subsection{Energy relaxation due to electron-phonon interaction}

Supercollisions are scattering events involving electron scattering
off a phonon and an impurity. As such, this is a next-order process as
compared to the direct electron-phonon scattering. The reason
supercollisions might be important is that the speed of sound is much
smaller than $v_g$. At high enough temperatures \cite{meig1,srl}
supercollisions indeed dominate, but at lower temperatures the direct
electron-phonon scattering cannot be neglected.

Energy relaxation and quasiparticle recombination due to
electron-phonon scattering was considered in Ref.~\cite{hydro0} within
the linear response theory. Since the macroscopic equations of the
linear response theory coincide with the linearized hydrodynamic
equations \cite{me1}, we can directly incorporate the corresponding
decay terms into our hydrodynamic theory. These decay terms appear in
Eq.~(\ref{ts}) through the matrix $\widehat{T}_S$. The model of
electron-phonon interaction introduced in Ref.~\cite{hydro0}
corresponds to the following choice of this matrix
\begin{equation}
\widehat{T}_{ep}=- \frac{1}{|\Delta|}
\begin{pmatrix}
\frac{\gamma}{ \tau_{Ec}}+\frac{1}{\tau_{Eb} } 
& -\frac{\gamma^2}{\mathcal{N}_2\tau_{Eb}}-\frac{\gamma}{\tau_{Ec}}
\cr 
-\frac{\gamma^2\mathcal{N}_2}{\gamma\tau_{Ec}}-\frac{\mathcal{N}_2}{\tau_{Ic}}-\frac{1}{\tau_{Eb}} 
& \frac{2\gamma}{\tau_{Ec}}+\frac{\gamma^2}{\mathcal{N}_2\tau_{Eb}}+\frac{\mathcal{N}_2}{\tau_{Ic}}
\end{pmatrix}\!,
\end{equation}
where
\[
\gamma_=\frac{\pi^2}{12\ln^22},
\quad 
\mathcal{N}_2=\frac{9\zeta(3)}{8\ln^32},
\quad 
\Delta=\gamma^2-\mathcal{N}_2,
\]
and $\tau_{Eb}\ll \tau_{Ec}\leq \tau_{Ic}$ describe the three
independent components of the electron-phonon collision integral
\cite{hydro0}.

\begin{figure}[t]
\centering
\includegraphics[width=\columnwidth]{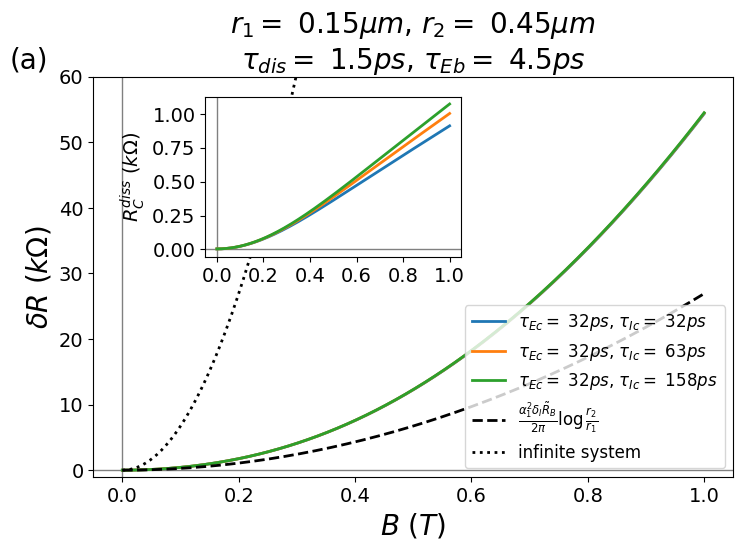}
\includegraphics[width=\columnwidth]{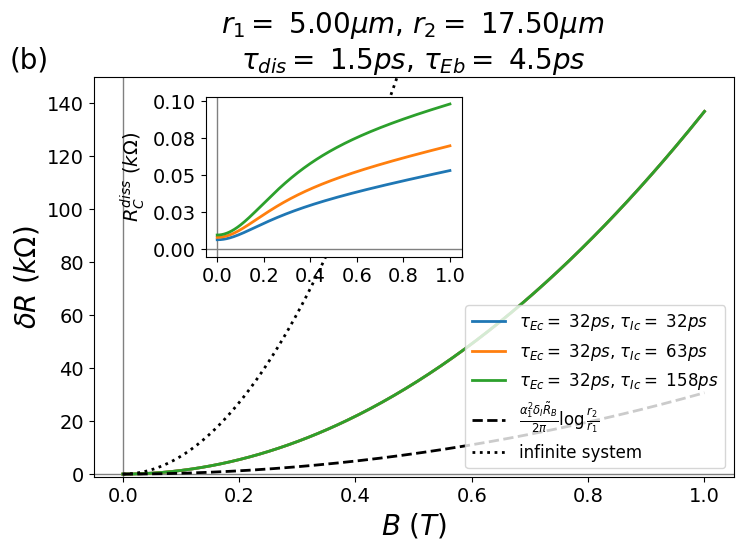}
\caption{Magnetoresistance in small (top) and large (bottom) Corbino
  devices computed within the ``electron-phonon model'' of energy
  relaxation (cf. Fig.~\ref{fig:rs}). }
\label{fig:rseph}
\end{figure}

\begin{figure}[t]
\centering
\includegraphics[width=\columnwidth]{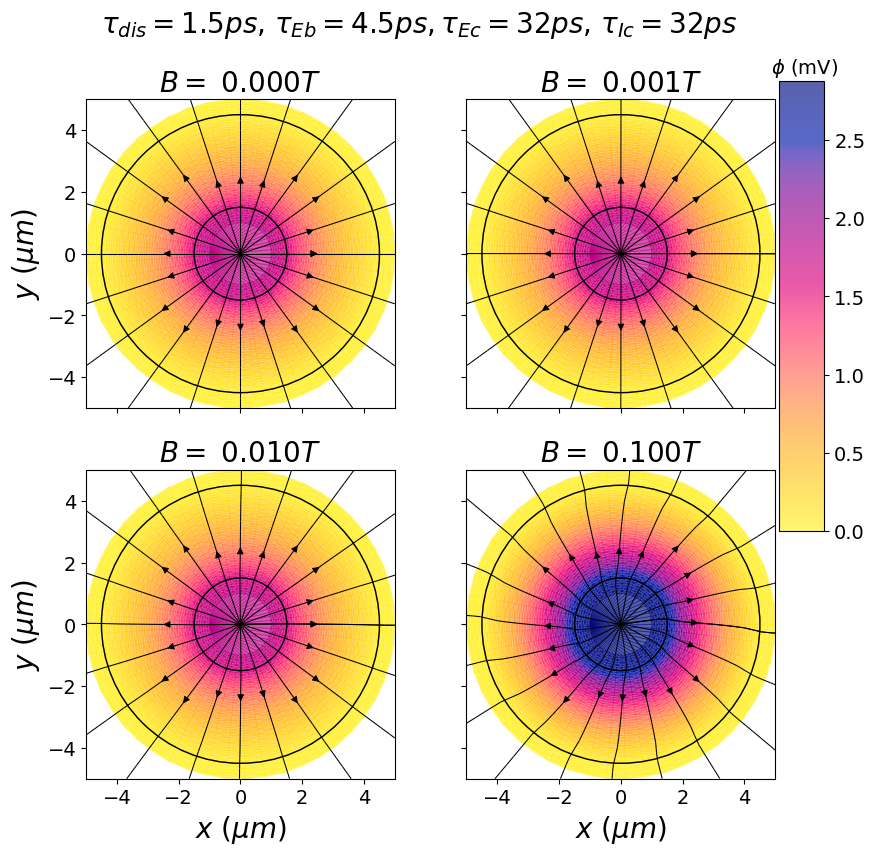}
\caption{Electric current density $\bs{j}$ and potential $\varphi$
  within the electron-phonon model of energy relaxation
  (cf. Fig.~\ref{fig:jf}).}
\label{fig:jfeph}
\end{figure}

\begin{figure}[t]
\centering
\includegraphics[width=\columnwidth]{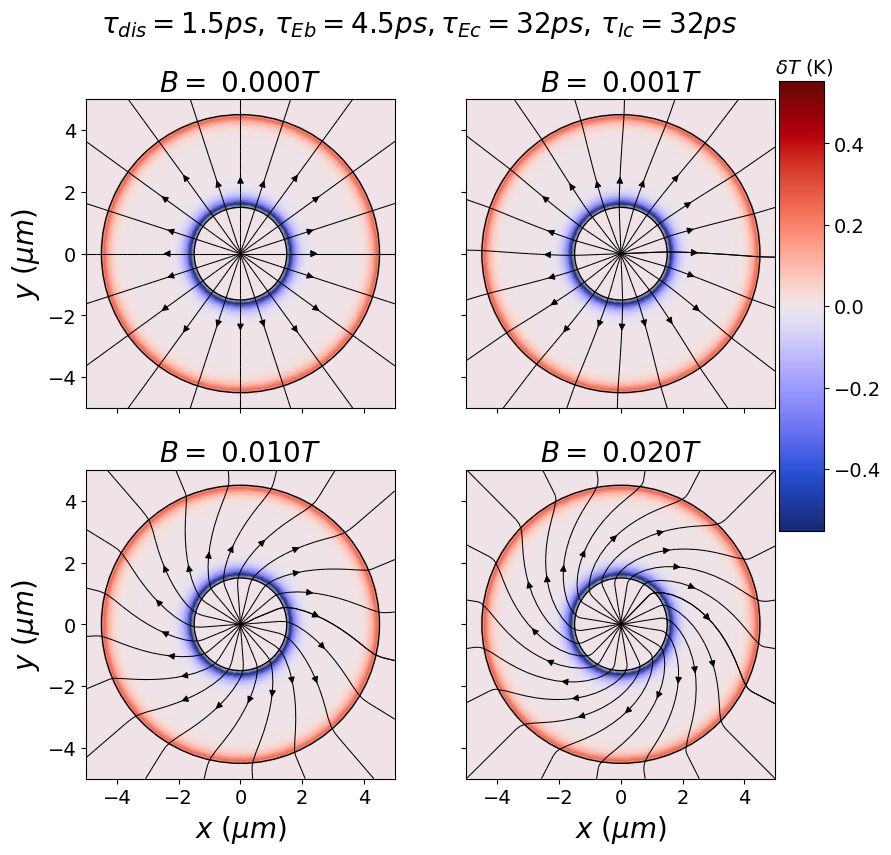}
\caption{Hydrodynamic velocity $\bs{u}$ and temperature $\delta T$
  distribution in the device obtained by solving the hydrodynamic
  equations at relatively low temperatures where energy relaxation is
  dominated by direct electron-phonon scattering
  (cf. Fig.~\ref{fig:ut}).}
\label{fig:uteph}
\end{figure}

\begin{figure}[t]
\centering
\includegraphics[width=\columnwidth]{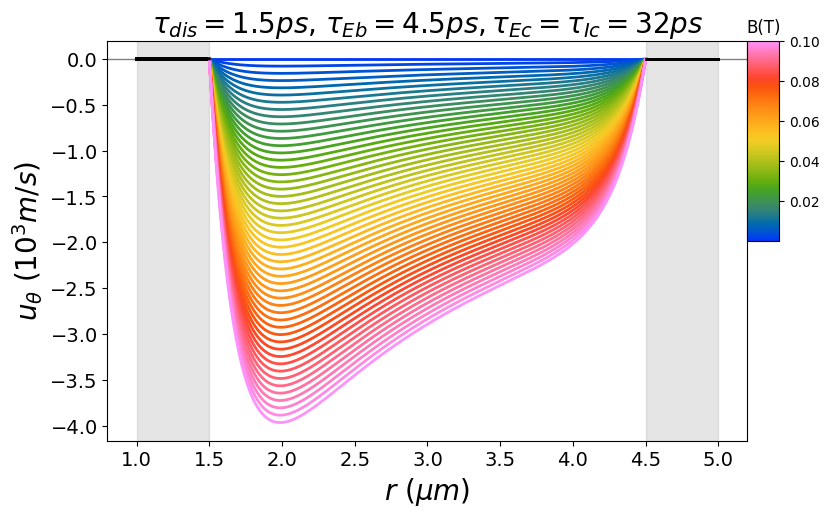}
\caption{Tangential component of the hydrodynamic velocity
  $u_\vartheta$ computed within the ``electron-phonon model'' of
  energy relaxation (cf. Fig.~\ref{fig:us}).}
\label{fig:uteph2}
\end{figure}

\begin{figure}[t]
\centering
\includegraphics[width=0.98\columnwidth]{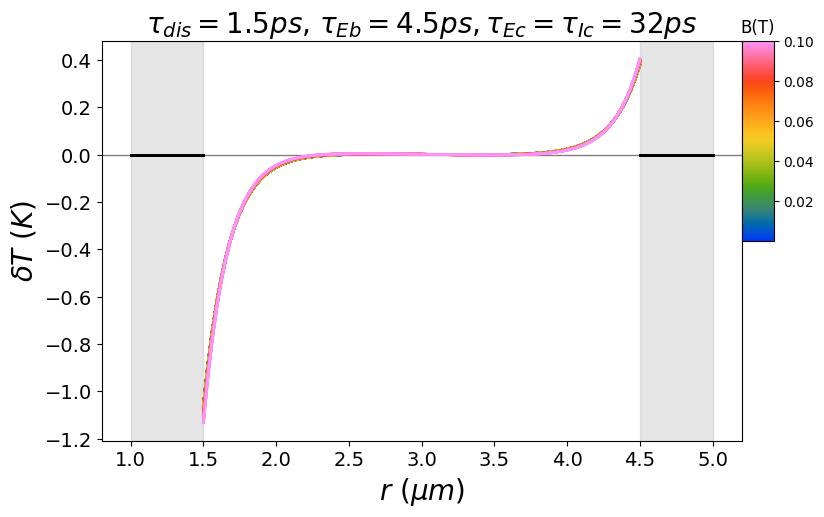}
\caption{Local temperature variation computed within the
  ``electron-phonon model'' of energy relaxation (cf.
  Fig.~\ref{fig:us}).}
\label{fig:teph}
\end{figure}

\begin{figure}[t]
\centering
\includegraphics[width=0.98\columnwidth]{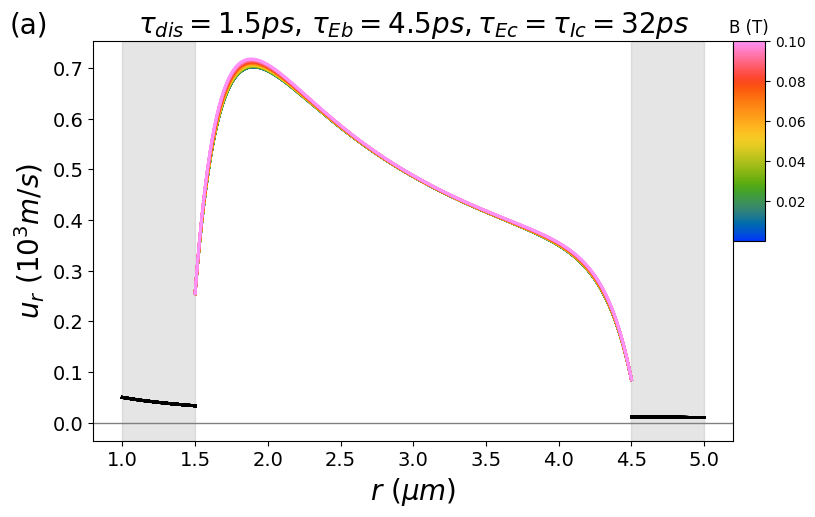}
\includegraphics[width=0.98\columnwidth]{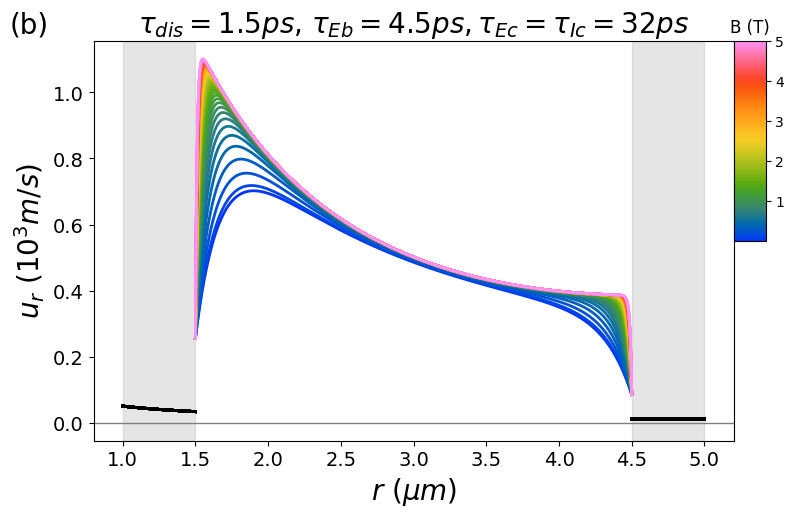}
\caption{Radial component of the hydrodynamic velocity $u_r$ computed
  within the ``electron-phonon model'' of energy relaxation
  (cf. Fig.~\ref{fig:us}).}
\label{fig:ureph}
\end{figure}

Repeating the above calculation with $\widehat{T}_{ep}$ instead of
$\widehat{T}_S$, we arrive at the results that are largely similar to
those obtained within the supercollision model, but with a few notable
differences (see Figs.~\ref{fig:rseph}-\ref{fig:ureph}). Unless
specified in the figure captions, the parameter values used for the
quantitative computation are the same as in the case of
supercollisions (see the caption to Fig.~\ref{fig:ut}).

Magnetoresistance of the device is still positive and parabolic (see
Fig.~\ref{fig:rseph}). In small devices, it is still largely determined
by the quantity $\tilde R$ (shown by the black dotted line in
Fig.~\ref{fig:rseph} similarly to Fig.~\ref{fig:rs}). In this case,
variations of the electron-phonon relaxation rates still do not affect
the result in any noticeable way. The results for large devices are
also similar to the case of supercollisions: calculated
magnetoresistance clearly exceeds $\tilde R$ and thus shows a strong
dependence on the size of the device (but not on the ratio $p$).

The electric current density and potential in the device are seen
largely the same as in the case of supercollisions, although the
deviation of the current from the radial direction (i.e., its
tangential component $\delta j_\vartheta$) is somewhat smaller (see
Fig.~\ref{fig:jfeph}, cf. Fig.~\ref{fig:jf}). This result seems to be
consistent with the similarities in the magnetoresistance in the two
cases.

The hydrodynamic velocity $\bs{u}$ is still dominated by its
tangential component (see Figs.~\ref{fig:uteph} and
\ref{fig:uteph2}). The latter shows the behavior that is largely
similar to that shown in the bottom panel of Fig.~\ref{fig:us},
although the magnitude of $u_\vartheta$ shows stronger growth with
increasing magnetic field. In contrast, the temperature variation is
``reversed'': now the electronic temperature is increased around the
inner contact and decreased close to the outer one (the opposite
behavior to that seen in Figs.~\ref{fig:ut} and \ref{fig:tps}) (see
Fig.~\ref{fig:teph}).


The reversed behavior of the temperature variation corresponds to the
change in the radial component of the hydrodynamic velocity
$u_r$. While the jumps at the interfaces with the leads remain the
same (insofar $u_r$ on the sample side of the interface is larger than
the drift velocity in the leads), the initial slope of $u_r$ as a
function of the radial coordinate has the opposite sign, which
does not change with the increase in the magnetic field.

Overall, it is rather natural that the choice of the energy relaxation
model mostly affects the energy flow in the device rather than the
charge flow. This is a clear consequence of the decoupling of the
energy and electric currents in neutral graphene. Although the two
currents are being coupled by the magnetic field, the effect appears
to be subleading. It is not surprising that the effect of this
coupling is most pronounced in strong magnetic fields and large
Corbino disks.

Contact resistance induced by viscous dissipation (see insets in
Figs.~\ref{fig:rs} and \ref{fig:rseph}) is also affected by the choice
of the energy relaxation model. In the case of supercollisions its
qualitative behavior exhibits a strong dependence on the size of the
disk (see Fig.~\ref{fig:rs}), while in the model of electron-phonon
scattering this dependence is reduced to the magnitude only. The
contact resistance is significantly stronger in small devices for both
choices of the energy relaxation model as expected on general grounds.

\section{Summary}

In this paper we considered hydrodynamic flows of charge and energy in
neutral graphene Corbino disks. We have shown that the Corbino
geometry offers a (in principle realizable) possibility to measure
electronic viscosity in neutral graphene, a task that so far has
appeared elusive. The viscosity coefficient could be extracted from
the magnetoresistance data in the ultra-clean limit where the bulk
contribution to the device resistance is independent of the
electron-impurity scattering time. The bulk resistance dominates over
the contact resistance for larger sized disks and hence can in
principle be measured in laboratory experiments.

Corbino magnetoresistance in graphene is illustrated in
Figs.~\ref{fig:rs} and \ref{fig:rseph}, where the calculated
magnetoresistance is shown for two models of energy relaxation. In
both cases, the dependence $R(B)$ is parabolic, similarly to the known
result in the strip geometry. The viscosity coefficient can be {\it in
  principle} determined experimentally by analyzing the data in a set
of different Corbino disks (see Sec.~\ref{sec_rb}). This is not a
straightforward task since the magnetoresistance is given by a sum of
viscosity-dependent and viscosity-independent terms.  In the clean
limit $\ell_G\ll r_1, r_2$ [see Eq.~(\ref{ub1})], these terms exhibit
distinct dependence on the sample size $r_2$, the ratio of the radii
$p=r_2/r_1$, and temperature, making it possible to extract the
viscosity coefficient from the experimental data. In the opposite
limit [see Eq.~(\ref{ub2})], the dominant contribution to
magnetoresistance is independent of viscosity. Existing experiments
appear to be in the crossover between these two limits. In this paper
we have used parameter values yielding $\ell_G\approx0.2\,\mu$m. The
size of the Corbino disk used in a recent experiment \cite{sulp22} was
$r_1=2\,\mu$m, $r_2=9\,\mu$m, which is closer to the ``large Corbino
disk'' illustrated in panels (b) in Figs.~\ref{fig:rs} and
\ref{fig:rseph} than to the clean limit. It is fair to say that at
present extracting viscosity from Corbino magnetoresistance
measurements would be extremely difficult. At the same time, we are
not aware of any other way to measure the viscosity coefficient in
neutral graphene. We believe that viscosity measurements and more
generally experimental observation of purely viscous effects in
neutral graphene will be more accessible in the near future with even
cleaner samples (increasing $\tau_{\rm dis}$ by an order of
magnitude).

The regime of linear magnetoresistance as seen
in the strip geometry or infinitely sized models does not exist in the
Corbino geometry. This can be easily understood by noting that the
origin of linear magnetoresistance is in the accumulation of energy
and quasiparticle density in the boundary region of a long strip where
the sample edges provide a natural barrier for the lateral neutral
flow of quasiparticles induced by the magnetic field. In a Corbino
disk there is no such edge. The lateral currents (energy and
imbalance) flow freely around the disk without accumulating
quasiparticles at any point.

Unlike the case of a single-band conductor (e.g., doped graphene), at
charge neutrality the electric field is not expelled from the bulk of
the sample. Nevertheless bulk viscous dissipation does lead to a
discontinuity of the electric potential at the sample-lead interfaces
inducing an additional contact resistance. This resistance however is
rather small as compared to the resistance of the whole device and
should not have a strong effect on the viscosity measurements.

\section*{Acknowledgments} 

The authors are grateful to P. Hakonen, V. Kachorovskii, A. Levchenko,
A. Mirlin, J. Schmalian, A. Shnirman, and M. Titov for fruitful
discussions. This work was supported by the German Research Foundation
DFG within FLAG-ERA Joint Transnational Call (Project GRANSPORT), by
the European Commission under the EU Horizon 2020 MSCA-RISE-2019
Program (Project No. 873028 HYDROTRONICS), by the German Research
Foundation DFG project NA 1114/5-1 (B.N.N.), and by the German-Israeli
Foundation for Scientific Research and Development (GIF) Grant
No. I-1505-303.10/2019 (I.V.G.).

\bibliography{hydro-refs,refs-books}
\end{document}